\pdfoutput=1

\documentclass[11pt]{article}

\usepackage{acl}

\usepackage{graphicx}
\graphicspath{ {./img/} }

\usepackage{times}
\usepackage{latexsym}
\usepackage{amsmath}
\usepackage{bm}
\usepackage{tabu}
\usepackage{tablefootnote}

\usepackage[T1]{fontenc}

\usepackage[utf8]{inputenc}

\usepackage{microtype}

\title{PaddleSpeech: An Easy-to-Use All-in-One Speech Toolkit}

\author{Hui Zhang$^1$, Tian Yuan$^1$, Junkun Chen$^{3}$, Xintong Li$^2$, Renjie Zheng$^2$,  \\
    {\bf Yuxin Huang$^1$,  Xiaojie Chen$^1$,  Enlei Gong$^1$, Zeyu Chen$^1$, Xiaoguang Hu$^1$, } \\
    {\bf Dianhai Yu$^1$, Yanjun Ma$^1$, Liang Huang$^{2,3}$} \\
    $^1${Baidu Inc., Beijing, China}\\
    $^2${Baidu Research, Sunnyvale, CA, USA}\\
    $^3${Oregon State University, Corvallis, OR, USA}\\
    \texttt {\{zhanghui41, yuantian01, xintongli, renjiezheng,}\\
    \texttt {
    huangyuxin, chenxiaojie06, gongenlei, chenzeyu01\}\@baidu.com} 
}

\usepackage{hyperref}
\usepackage{xurl}
\usepackage[hang,flushmargin]{footmisc}

\usepackage{booktabs}
\usepackage{diagbox,tabularx}
\usepackage{graphicx}

\usepackage{fontawesome}
\usepackage{bm}
\usepackage{dsfont}
\usepackage{amsmath}
\usepackage{amssymb}
\usepackage{amsthm}
\usepackage{algorithm}
\usepackage{xcolor,colortbl}
\usepackage{subfigure}
\usepackage{xspace}
\usepackage{CJKutf8}
\usepackage{makecell}
\usepackage{multirow}
\usepackage{threeparttable}

\usepackage{lipsum}

\usepackage{tikz}
\usetikzlibrary{arrows.meta}
\usetikzlibrary{positioning}
\usetikzlibrary{calc}
\usetikzlibrary{shapes.symbols}
\usetikzlibrary{backgrounds}
\usetikzlibrary{fit}

\begin{document}
\begin{CJK}{UTF8}{gbsn}
\maketitle
\begin{abstract}

PaddleSpeech is an open-source all-in-one speech toolkit.
It aims at facilitating the development and research of speech processing technologies by providing an easy-to-use command-line interface and a simple code structure.
This paper describes the design philosophy and core architecture of PaddleSpeech to support several essential speech-to-text and text-to-speech tasks.
PaddleSpeech achieves competitive or state-of-the-art performance on various speech datasets and implements the most popular methods.
It also provides recipes and pretrained models to quickly reproduce the experimental results in this paper.
PaddleSpeech is publicly avaiable at \url{https://github.com/PaddlePaddle/PaddleSpeech}.%
\footnote{Demo video: \url{https://paddlespeech.readthedocs.io/en/latest/demo_video.html}}
\end{abstract}

\begin{table*}[t]
\small
\setlength{\tabcolsep}{2pt}
\centering
\begin{tabular}{llll}
\toprule
\textbf{Task} & \textbf{Description} & \textbf{Techniques} & \textbf{Datasets} \\
\midrule
\makecell[l]{Sound\\Classification} & \textit{Label sound class} & \makecell[l]{Finetuned PANN \cite{Kong2020PANNsLP}} & \makecell[l]{ESC-50 dataset \cite{Piczak2015ESCDF}}\\
\midrule
\makecell[l]{Speech\\Recognition} & \makecell[l]{\textit{Transcribe speech}\\\textit{to text}} & \makecell[l]{Deepspeech2 \cite{amodei2016deep}\\Conformer \cite{zhang2020unified}\\Transformer \cite{zhang2020unified}} & \makecell[l]{Librispeech \citep{panayotov2015librispeech}\\AISHELL-1 \citep{bu2017aishell}
}\\
\midrule
\makecell[l]{Punctuation\\Restoration} & \makecell[l]{\textit{Post-add punctuation}\\\textit{to transcribed text}} & \makecell[l]{Finetuned ERNIE \citep{Sun2019ERNIEER}} & \makecell[l]{IWSLT2012-zh \citep{federico2012overview}}\\
\midrule
\makecell[l]{Speech\\Translation} & \makecell[l]{\textit{Translate speech}\\\textit{to text}} & \makecell[l]{Transformer \citep{vaswani2017attention}} & \makecell[l]{MuST-C \cite{di2019must}}\\
\midrule
\makecell[l]{Text\\To Speech} & \makecell[l]{\textit{Synthesis speech}\\\textit{from text}} & \makecell[l]{\textbf{Acoustic Model}\\~~Tacotron 2 \citep{shen2018natural}\\~~Transformer  TTS \citep{li2019neural}\\~~SpeedySpeech \citep{vainer2020speedyspeech}\\~~FastPitch \citep{lancucki2021fastpitch}\\~~FastSpeech 2    \citep{ren2020fastspeech}\\\textbf{Vocoder}\\~~WaveFlow\cite{ping2020waveflow}\\~~Parallel WaveGAN \citep{yamamoto2020parallel}\\~~MelGAN \citep{kumar2019melgan}\\~~Style MelGAN \citep{mustafa2021stylemelgan}\\~~Multi Band MelGAN \citep{yang2021multi}\\~~HiFi GAN \citep{kong2020hifi}} & \makecell[l]{CSMS (DataBaker)\\AISHELL-3 \citep{shi2020aishell}\\LJSpeech \citep{ljspeech17}\\VCTK \citep{yamagishi2019cstr}}\\
\bottomrule
\end{tabular}
\caption{List of speech tasks and corpora that are currently supported by PaddleSpeech.}
\label{tab:intro}
\end{table*}

\section{Introduction}

Speech processing technology enables humans to directly communicate with computers, which is an essential part of enormous applications such as smart home devices \cite{hoy2018alexa}, autonomous driving, and simultaneous translation \cite{zheng2020fluent}. 
Open-source toolkits boost the development of speech processing technology by lowering the barrier of application and research in this area \citep{young2002htk, lee2001julius, huggins2006pocketsphinx, rybach2011rasr, povey2011kaldi, watanabe2018espnet, han2019delta, wang-etal-2020-fairseq, ravanelli2021speechbrain, zhao-etal-2021-neurst}.

However, the current prevailing speech processing toolkits presume that their users are experienced practitioners or researchers, so beginners might feel baffled when developing their exciting applications.
For example, to prototype new speech applications with Kaldi \citep{povey2011kaldi}, the users have to be comfortable reading and revising the provided recipes written in Bash, Perl, and Python scripts and be proficient at C++ to hack its implementation.
The more recent toolkits, such as Fairseq S2T \citep{wang-etal-2020-fairseq} and NeurST \citep{zhao-etal-2021-neurst}, become more flexible by building on general-purpose deep learning libraries.
But their complicated code styles also make it time-consuming to learn and hard to migrate from one to another.
So, we have developed PaddleSpeech, providing a command-line interface and portable functions to make the development of speech-related applications accessible to everyone.

Notably, the Chinese community has many developers eager to contribute to the community. However, nearly all deep learning libraries, such as Pytorch \citep{paszke2019pytorch} and Tensorflow \citep{abadi2016tensorflow}, target the English community mainly, so it significantly increases the difficulty for Chinese developers.
PaddlePaddle, as the only fully-functioning open-source deep learning platform targeting both the English and Chinese community, has accumulated more than 500k commits, 476k models, and is used by 157k enterprises.
So, we expect PaddleSpeech, developed with PaddlePaddle can remove the barriers between the English and Chinese communities to boost the development of speech technologies and applications.

Developing speech applications for the industry is not the same scenario as conducting research in academia.
The research papers mainly focus on developing novel models to perform better on specific datasets.
However, a clean dataset usually does not exist when applying a speech product.
So, PaddleSpeech provides on-the-fly preprocessing for the raw audios to make PaddleSpeech directly usable in product-oriented applications.
Notably, some preprocessing methods are exclusive in PaddleSpeech, such as rule-based Chinese text-to-speech frontend, which can significantly benefit the performance of synthesized speech.

Performance is the cornerstone of all applications.
PaddleSpeech achieves state-of-the-art or competitive performers on various commonly used benchmarks, as shown in Table \ref{tab:intro}.

Our main contributions in this paper are two-folds.
\begin{itemize}
    \item We introduce how we designed PaddleSpeech and what features it supports.
    \item We provide the implementation and reproducible experimental details that result in state-of-the-art or competitive performance on various tasks.
\end{itemize}

\begin{figure}[t]
    \centering
    \resizebox{\columnwidth}{!}{
        \begin{tikzpicture}
\small

\tikzstyle{box} = [rectangle, draw=black!100, anchor=south west];
\tikzstyle{Bigbox} = [thick, fill=blue!10, rounded corners, rectangle, dashed, minimum width=8.5cm];

\node[box, minimum width=5cm, minimum height=1cm] (paddle) {PaddlePaddle};
\node[right=.3cm of paddle.south east, box, minimum width=3cm, minimum height=1cm] (ol) {};
\node[anchor=north west] at (ol.north west) {Other Libraries};
\node[anchor=center, yshift=-.15cm] at (ol.center) {\tiny Kaldi, Sclite, Sox, Openblas};
\begin{pgfonlayer}{background}
    \node[Bigbox, draw=blue!50, fill=blue!10, yshift=-.2cm, minimum height=1.65cm] [fit = ($(paddle.west)!.5!(ol.east)$) ] (layer0) {};
    \node[anchor=south] at (layer0.south) {\textit{Fundamental Platforms}};
\end{pgfonlayer}

\node[above=.8cm of paddle.north west, box, minimum width=5.5cm, minimum height=1cm] (audio) {};
\node[anchor=north west] at (audio.north west) {PaddleAudio};
\node[anchor=center, yshift=-.15cm] at (audio.center) {\tiny \makecell[l]{paddleaudio.features.Spectrogram\\paddleaudio.features.LogMelSpectrogram}};
\node[right=.3cm of audio.south east, box, minimum width=2.5cm, minimum height=1cm] (util) {};
\node[anchor=north west] at (util.north west) {Utils};
\node[anchor=center, yshift=-.15cm] at (util.center) {\tiny \makecell[l]{utils/build\_vocab.py\\utils/parse\_options.sh}};
\begin{pgfonlayer}{background}
    \node[Bigbox, draw=green!50, fill=green!10, yshift=-.2cm, minimum height=1.65cm] [fit = ($(audio.west)!.5!(util.east)$) ] (layer1) {};
    \node[anchor=south] at (layer1.south) {\textit{Common Modules}};
\end{pgfonlayer}

\node[above=.8cm of audio.north west, box, minimum width=5cm, minimum height=1.5cm] (speech) {};
\node[anchor=north west] at (speech.north west) {PaddleSpeech};
\node[anchor=center, yshift=-.15cm] at (speech.center) {\tiny \makecell[l]{cli/asr/infer.py\\t2s/exps/fastspeech2/train.py\\t2s/models/fastspeech2/fastspeech2.py\\t2s/models/fastspeech2/fastspeech2\_updater.py}};
\node[right=.3cm of speech.south east, box, minimum width=3cm, minimum height=1.5cm] (data) {};
\node[anchor=north west] at (data.north west) {Datasets};
\node[anchor=center, yshift=-.15cm] at (data.center) {\tiny \makecell[l]{dataset/aishell/aishell.py\\dataset/librispeech/librispeech.py}};
\begin{pgfonlayer}{background}
    \node[Bigbox, draw=yellow!50, fill=yellow!10, yshift=-.2cm, minimum height=2.15cm] [fit = ($(speech.west)!.5!(data.east)$) ] (layer2) {};
    \node[anchor=south] at (layer2.south) {\textit{Models \& Updaters \& Datasets}};
\end{pgfonlayer}

\node[above=.8cm of speech.north west, box, minimum width=8.3cm, minimum height=1cm] (demo) {};
\node[anchor=north west] at (demo.north west) {Examples/Demos};
\node[anchor=center, yshift=-.15cm] at (demo.center) {\tiny \makecell[l]{examples/aishell/run.sh\\demos/metaverse/run.sh}};
\begin{pgfonlayer}{background}
    \node[Bigbox, draw=red!50, fill=red!10, yshift=-.2cm, minimum height=1.65cm] [fit = ($(demo.west)!.5!(demo.east)$) ] (layer3) {};
    \node[anchor=south] at (layer3.south) {\textit{Recipes}};
\end{pgfonlayer}

\end{tikzpicture}
    }
    \caption{Software architecture of PaddleSpeech.}
    \label{fig:layout}
\end{figure}
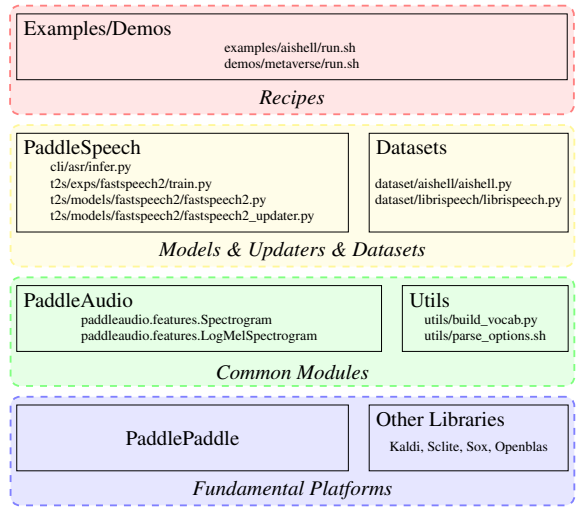




\section{Design of PaddleSpeech}

Figure~\ref{fig:layout} shows the software architecture of PaddleSpeech.
As an easy-to-use speech processing toolkit, PaddleSpeech provides many complete recipes to perform various speech-related tasks and demo usage of the command line interface.
Getting familiar with \colorbox{red!10}{the top level} should be enough for building speech-related applications.

\colorbox{yellow!10}{The second level} faces researchers in speech and language processing.
The design philosophy of PaddleSpeech is model-centric to simplify the learning and development of speech processing methods.
For a specific method, all computations of a specific model are included in two files under {\small \texttt{PaddleSpeech/<task>/models/<model>}}.%
\footnote{\texttt{<task>} includes \texttt{s2t} and \texttt{t2s} which stands for speech-to-text and text-to-speech respectively.}


PaddleSpeech has implemented most of the commonly used and well-performing models.
A model architecture is implemented in a standalone file named by the method.
Its corresponding training step and evaluation step are implemented in another {\small \texttt{updater}} file.
Generally, reading or hacking these two files is enough to understand or design a model.
More advanced hacking on more fine data processing or more complicated training/evaluation loop is also available at {\small \texttt{PaddleSpeech/<task>/exps/<model>}}.
The original datasets can be obtained by scripts in corresponding {\small \texttt{dataset/<dataset>/}}.
PaddleSpeech supports distributed multi-GPU training with good efficiency.

The standard modules, such as audio and text feature transformation and utility scripts, are implemented as libraries in \colorbox{green!10}{the third level}.
The backend of PaddleSpeech is mainly PaddlePaddle with some functions from third-party libraries as shown in \colorbox{blue!10}{the fourth level}.
PaddleSpeech provides multiple ways to extract multiple types of speech features from raw audios using PaddleAudio and Kaldi, such as spectrogram and filterbanks, which can be varied according to the needs of the tasks.

\section{Experiments}

In this section, we compare the performance of models in PaddleSpeech with other popular implementations in five speech-related tasks, including sound classification, speech recognition, punctuation, speech translation, and speech synthesizing. 
The toolkit can reach SOTA on most tasks.
All experiments in this section include details on data preparation, evaluation metrics, and implementation to enhance reproducibility.%
\footnote{\url{https://github.com/PaddlePaddle/PaddleSpeech/tree/develop/examples}}

\subsection{Sound Classification}
Sound Classification is a task to recognize particular sounds, including speech commands \citep{Warden2018SpeechCA}, environment sounds \cite{Piczak2015ESCDF}, identifying musical instruments \cite{Engel2017NeuralAS}, finding birdsongs \cite{Stowell2018AutomaticAD}, emotion recognition \cite{xu2019alignemotion} and speaker verification \cite{liu2018robustsv}.

\paragraph{Datasets}
In this section, we analyze the performance of PaddleSpeech in Sound Classification on ESC-50 dataset \citep{Piczak2015ESCDF}.
The ESC-50 dataset is a labeled collection of 2000 environmental 5-second audio recordings consisting of 50 sound events, such as "Dog", "Cat", "Breathing" and "Fireworks", with 40 recordings per event.

\paragraph{Data Preprocessing}
First, we resample all audio recordings to 32 kHz, and convert them to monophonic to be consistent with the PANNs trained on AudioSet \citep{Kong2020PANNsLP}. 
And then, we transform the recordings into log mel spectrograms by applying
short-time Fourier transform on the waveforms with a Hamming window of size 1024 and a hop size of 320 samples.
This configuration leads to 100 frames per second.
Following \citet{Kong2019CrosstaskLF}, we apply 64 mel filter banks to calculate the log mel spectrogram.

\paragraph{Implementation}
PANNs \cite{Kong2020PANNsLP} is one of the pre-trained CNN models for audio-related tasks, which is characterized in terms of being trained with the AudioSet \cite{Gemmeke2017AudioSA}.
PANNs are helpful for tasks where only a limited number of training clips are provided.
In this case, we fine-tune all parameters of a PANN for the environment sounds classification task.
All parameters are initialized from the PANN, except the final fully-connected layer which is randomly initialized.
Specifically, we implement CNNs with 6, 10 and 14 layers, respectively  \cite{Kong2020PANNsLP}.

\paragraph{Results}
We report 5-fold cross validation accuracy values on ESC-50 dataset.
As shown in Table \ref{tab:esc50}, PANNs-CNN14 achieves 0.9500 5-fold cross validation accuracy that is comparable to the current state-of-the-art method \cite{Gong2021ASTAS}.

\begin{table}
\small
\begin{center}
\begin{tabular}{ll}
\toprule
Model                       & Accuracy \\
\midrule
AST-P \citep{Gong2021ASTAS} & $95.6 \pm 0.4$ \\
\midrule
PANNs-CNN14                 & 95.00   \\
PANNs-CNN10                 & 89.75   \\
PANNs-CNN6                  & 88.25   \\
\bottomrule
\end{tabular}
\caption{5-fold cross validation accuracy of ESC-50.}
\label{tab:esc50}
\end{center}
\end{table}

\begin{table*}[t]
\small
\centering
\begin{threeparttable}
\begin{tabular}{clcclll}
\toprule
Data & Model & Streaming & Test Data & Language Model & CER & WER \\
\midrule
\multirow{5}{*}{\rotatebox{90}{Aishell}}
& WeNet Conformer$^\dagger$$^*$ \cite{yao2021wenet}
& \checkmark & & & 5.45 & - \\
& WeNet Conformer$^\dagger$ \cite{yao2021wenet}
& & & & 4.61 & - \\
& WeNet Transformer$^\dagger$ \citep{yao2021wenet} & & & & 5.30 & -\\
& ESPnet Conformer$^\dagger$ \citep{inaguma2020espnet} &  & &  & 5.10 & - \\
& ESPnet Transformer$^\dagger$ \citep{inaguma2020espnet} &  & &  & 6.70 & - \\
& SpeechBrain Transformer$^\dagger$ \citep{ravanelli2021speechbrain} & & & & 5.58 & -\\

\cmidrule{2-7}
& Deepspeech 2 & \checkmark & & char 5-gram
& 6.66 & -  \\
& Deepspeech 2 & & & char 5-gram
& 6.40 & - \\
& Transformer & & & & 5.23 & -\\
& Conformer$^*$ & \checkmark &  & & 5.44 & -  \\
& Conformer & &  & & 4.64 & -  \\

\midrule
\multirow{6}{*}{\rotatebox{90}{Librispeech}}
& WeNet Conformer$^\dagger$ \cite{yao2021wenet}& & test-clean & & - & 2.85 \\
& SpeechBrain Transformer$^\dagger$ \citep{ravanelli2021speechbrain} &  & test-clean & TransformerLM & - & 2.46 \\
& ESPnet Transformer$^\dagger$ \citep{inaguma2020espnet} &  & test-clean & TransformerLM
& - & 2.60 \\
\cmidrule{2-7}
& Deepspeech 2 & & test-clean & word 5-gram
& - & 7.25  \\
& Conformer & & test-clean & & - & 3.37  \\
& Transformer & & test-clean & TransformerLM
& - & 2.40 \\
\bottomrule
\end{tabular}
\begin{tablenotes}
    \scriptsize
    \item[$^\dagger$] denotes the results are reported in their public repositories.
    \item[$^*$] denotes the results are streaming with chunk size 16.
\end{tablenotes}
\end{threeparttable}
\caption{WER/CER on Aishell, Librispeech for ASR Tasks.}
\label{tab:asr-result}
\end{table*}

\subsection{Automatic Speech Recognition}

Automatic Speech Recognition (ASR) is a task to transcribe the audio content to text in the same language.

\paragraph{Datasets}
We conduct the ASR experiments on two major datasets including
Librispeech\footnote{\url{http://www.openslr.org/12/}} \citep{panayotov2015librispeech}
and
Aishell-1\footnote{\url{http://www.aishelltech.com/kysjcp}} \citep{bu2017aishell}.
Librispeech contains 1000 hours speech data.
The whole dataset is divided into 3 training sets (100h clean, 360h clean, 500h other), 2 validation sets (clean, other), and 2 test sets (clean, other).
Aishell contains 178 hours speech data.
400 speakers from different accent areas in China participate in the recording.
The dataset is divided into the training set (340 speakers), validation set, (40 speakers) and test set (20 speakers).


\paragraph{Data Preprocessing}
Deepspeech 2 takes character-level vocabularies for both English and Mandarin tasks.
For other models, we use character-level vocabulary for Mandarin. 
And English text is preprocessed with SentencePiece \cite{kudo2018sentencepiece}.
Both two kinds of datasets are added four additional characters, which are \texttt{<'>}, \texttt{<space>}, \texttt{<blank>} and \texttt{<eos>}.
For cepstral mean and variance normalization (CMVN), a subset of or full of the training set is selected and be used to compute the feature mean and standard error. 
For feature extraction, we have several methods implemented, such as linear spectrogram, filterbank, and mfcc.
Currently, the Deepspeech 2 model uses linear spectrogram or filterbank, but Transformer and Conformer models use filterbank.
For a fair comparison, we take additional 3 dimensional pitch features into Transformer to be consistent with ESPnet.

\paragraph{Implementation}
We implement both streaming and non-streaming Deepspeech 2 \citep{amodei2016deep}.
The non-streaming model has 2 convolution layers and 3 LSTM layers. 
The streaming model has 2 convolution layers and 5 LSTM layers.
The Conformer and Transformer models are implemented following \citet{zhang2020unified} with 12 encoder layers and 6 decoder layers.

\paragraph{Results}
We report word error rate (WER) and character error rate (CER) for Librispeech (English) and Aishell (Mandarin) speech recognition, respectively.
As shown in Table \ref{tab:asr-result}, Conformer and Transformer are better than Deepspeech 2.
Our best models achieve comparable performance on both datasets compared with related works.

\begin{table*}[t]
\small
\centering
\begin{tabular}{lcccccccc}
\toprule
\textbf{Frameworks}                                    & \textbf{De}   & \textbf{Es}   & \textbf{Fr}   & \textbf{It}   & \textbf{Nl}   & \textbf{Pt}   & \textbf{Ro}   & \textbf{Ru}   \\
\midrule
ESPnet-ST \cite{inaguma2020espnet}       & 22.9 & \textbf{28.0} & 32.8 & \textbf{23.8} & \textbf{27.4} & 28.0 & 21.9 & \textbf{15.8} \\
fairseq-ST \cite{wang-etal-2020-fairseq} & 22.7 & 27.2 & 32.9 & 22.7 & 27.3 & 28.1 & 21.9 & 15.3 \\
NeurST \cite{zhao-etal-2021-neurst}      & 22.8 & 27.4 & \textbf{33.3} & 22.9 & 27.2 & 28.7 & \textbf{22.2} & 15.1 \\
\midrule
PaddleSpeech                             & \textbf{23.0} & 27.4 & 32.9 & 22.9 & 26.7 & \textbf{28.8} & \textbf{22.2} & 15.4 \\
\bottomrule
\end{tabular}
\caption{Case-sensitive detokenized BLEU scores on MuST-C \textit{tst-COMMON}.}
\label{tab:mustc}
\end{table*}

\subsection{Punctuation Restoration}

Punctuation restoration is a post-processing problem for ASR systems.
It is crucial to improve the readability of the transcribed text for the human reader and facilitate down-streaming NLP tasks.

\paragraph{Datasets}
We conduct experiments on IWSLT2012-zh\footnote{\url{https://hltc.cs.ust.hk/iwslt/}} dataset,
which contains 150k Chinese sentences with punctuation.
We select comma, period, and question marks as restore targets in this task, so we replace other punctuation with these three marks before training a model.
We split the data into training, validation and testing sets with 147k, 2k, and 1k samples, respectively.

\paragraph{Implementation}
We formulate the problem of punctuation restoration as a sequence labeling task with four target classes including \texttt{EMPTY}, \texttt{COMMA}, \texttt{PERIOD}, and \texttt{QUESTION} \citep{Nagy2021AutomaticPR}.
ERNIE \citep{Sun2019ERNIEER}, as a pretrained language model,
achieves new state-of-the-art results on five Chinese natural language processing tasks, including natural language inference, semantic similarity, named entity recognition, sentiment analysis, and question answering.
So, we finetune an ERNIE model for this task.
More specifically, all parameters are initialized from the ERNIE pretrained model, except the final shared fully-connected layer, which is randomly initialized.

\paragraph{Results}
We report F1-score values on IWSLT2012-zh dataset.
As shown in Table \ref{tab:iwslt2012-zh-pr}, our ERNIELinear model achieves 0.6331 overall F1-score, which is comparable with the previous work \cite{nagy2021automatic}.


\begin{table}[t]
\small
\setlength{\tabcolsep}{2pt}
\centering
\begin{threeparttable}
\begin{tabular}{lcccc}
\toprule
model & COMMA  & PERIOD & QUESTION & Overall \\
\midrule
BERTLinear$^\dagger$  & 0.4646 & 0.4227 & 0.7400 & 0.5424  \\
BERTBiLSTM$^\dagger$ & 0.5190 & 0.5707 & 0.8095 & 0.6330  \\
\midrule
ERNIELinear & 0.5142 & 0.5447 & 0.8406   & 0.6331  \\
\bottomrule
\end{tabular}
\begin{tablenotes}
    \scriptsize
    \item[$^\dagger$] denotes the results come from our reproduced models.
\end{tablenotes}
\end{threeparttable}
\caption{F1-score values on IWSLT2012-zh dataset.}
\label{tab:iwslt2012-zh-pr}
\end{table}

\subsection{Speech Translation}

Speech translation, 
where translating speech in a source language to text in another language,
is beneficial in human communications.

\paragraph{Datasets}
In this section, we analyze the performance of speech-to-text translation with PaddleSpeech on MuST-C dataset \citep{di2019must} with 8 different language translation pairs, which take the English speech as the source input. 

\paragraph{Implementation}
We process the raw audios with Kaldi \citep{povey2011kaldi} 
and extract 80-dimensional log-mel filterbanks stacked with 3-dimensional pitch feature
using a 25ms window size and a 10ms step size.
Text is firstly tokenized with Moses tokenizer%
\footnote{\url{https://github.com/moses-smt/mosesdecoder}} 
and then processed by SentencePiece \cite{kudo2018sentencepiece} with
a joint vocabulary whose size is 8K for each language pair.
We employ Transformer \citep{vaswani2017attention} as the base architecture for the speech translation experiments.
In detail, 
the Transformer model has 12 encoder layers that follow 2 layers of 2D convolution
with kernel size of 3 and stride size of 2,
and 6 decoder layers.
Each layer contains 4 attention heads with a size of 256.
The encoder is initialized from a pretrained ASR model.

\paragraph{Results}
We report detokenized case-sensitive BLEU%
\footnote{\url{https://github.com/mjpost/sacrebleu}}.
As shown in Table \ref{tab:mustc}, 
PaddleSpeech can achieve competitive results compared with other frameworks.

\begin{table*}[tb]
\small
\setlength{\tabcolsep}{0pt}
\centering
\resizebox{\textwidth}{!}{
\begin{tabular}{ccccccccccc}
\toprule
 & \textbf{Module} & \textbf{Result} \\
\midrule
\multirow{9}{*}{\rotatebox{90}{\textbf{PaddleSpeech}}} & & \textit{jīn ti\=an} &  sh\`i & & & zu\`i dī & w\=en d\`u & sh\`i &   \\
& \textit{Text} & 今天 & 是 & \textcolor{red}{2020/10/29} & ，& 最低& 温度& 是& \textcolor{blue}{-3°C} & 。  \\
&  {\small\it} & today & is  &    & & lowest & temperature &is& \\ \cmidrule{2-11}
 & & & & \textcolor{red}{\textit{ \`er l\'ing \`er l\'ing n\'ian sh\'i y\`ue \`er sh\'i j\v{i}u r\`i}} & & & & & \textcolor{blue}{\textit{l\'ing xi\`a \;\;\; s\=an \; d\`u}} \\
& \textbf{TN} & 今天 &是 &\textcolor{red}{二\;零\;\;二\;\;零 \; 年 \;\; 十 \;月 \;二\;十\;九 \;日} &，&最低&温度&是 &\textcolor{blue}{零下 \;\;\;\; 三 \;\; 度} &。\\
&  & & & \textcolor{red}{2\;\; 0\;\;\; 0\;\;\; 2\;\; year\; 10 {\scriptsize month} \;\; 29 \;\;\;day} & & & & & \textcolor{blue}{\textit{negative three degree}} \\ \cmidrule{2-11}
& \textbf{WS} & 今天 /& 是 /& 二零二零年 / 十月 / 二十九日 &，& /  最低&温度 /& 是 /& 零下 / 三度 &。\\ \cmidrule{2-11}
& \textbf{G2P} & { jin1 tian1} & { shi4} & { er4 ling2 er4 ling2 nian2 shi4 yue4 er4 shi2 jiu3 ri4} &  & { zui4 di1} & { wen1 du4} & { shi4} & { ling2 xia4 san1 du4} & \\ \midrule
\multicolumn{2}{c}{\textbf{ESPnet}} & jin1 tian1 &  shi4 & 2020/10/29 & & zui4 di1 & wen1 du4 & shi4 & -3°C  \\
\bottomrule
\end{tabular}
}
\caption{An example of the text preprocessing pipeline for Mandarin TTS
of PaddleSpeech and ESPnet. \textbf{TN} stands for the text normalization module, \textbf{WS} stands for the word segmentation module,
\textbf{G2P} stands for the grapheme-to-phoneme module.
The text normalization module for mandarin of ESPnet is not able to correctly 
handle dates (2020/10/29) and temperatures (-3°C).
}
\label{tab:preprocess}
\end{table*}

\subsection{Text-To-Speech}

A Text-To-Speech (TTS) system converts given language text into speech.
PaddleSpeech's TTS pipeline includes three steps.
We first convert the original text into the characters/phonemes through the text frontend module.
Then, through an Acoustic model, we convert the characters or phonemes into acoustic features, such as mel spectrogram,
Finally, we generate waveform from the acoustic features through a Vocoder.
In PaddleSpeech, the text frontend is a rule-based model inspired by expert knowledge.
The Acoustic models and Vocoders are trainable.







\paragraph{Datasets}
In PaddleSpeech, we mainly focus on Mandarin and English speech synthesis.
We have benchmarks on 
CSMSC\footnote{\url{https://www.data-baker.com/open_source.html}},
AISHELL-3\footnote{\url{http://www.aishelltech.com/aishell_3}},
LJSpeech\footnote{\url{https://keithito.com/LJ-Speech-Dataset/}},
VCTK\footnote{\url{https://datashare.ed.ac.uk/handle/10283/3443}}.
Due to the limit of space, we only list the experimental results on CSMSC,
which includes 12 hours speech audio corresponding to 10k sentences.

\paragraph{Text Frontend} 
A text frontend module is used 
to extract linguistic features, characters and phonemes from given text.
It mainly includes:
    Text Segmentation,
    Text Normalization (TN),
    Word Segmentation (WS),
    Part-of-Speech Tagging,
    Prosody Prediction and 
    Grapheme-to-Phoneme (G2P) (see Table~\ref{tab:preprocess}).


For Mandarin, our G2P system consists of a polyphone module, which uses pypinyin and g2pM, 
and a tone sandhi module which uses rules based on chinese word segmentations.
To the best of our knowledge,
our Mandarin text frontend system is the most complete one
compared with other publicly released works.

\paragraph{Data Preprocessing}
PaddleSpeech TTS uses the following modules for data preprocessing%
\footnote{%
\url{https://github.com/PaddlePaddle/PaddleSpeech/blob/develop/examples/csmsc/tts3/local/preprocess.sh}}%
:
First, we use Montreal-Forced-Aligner to get the duration for corresponding phonemes.
Second, we extract mel spectrograms as the features (additional pitch 
and energy features for Fastspeech 2).
Last, we conduct the statistical normalization for each feature. 

\paragraph{Acoustic Model}
Acoustic models can be mainly classified into autoregressive and non-autoregressive models.
The decoding of the autoregressive model relies on previous predictions at each step,
which leads to longer inference time but relatively better quality.
While the non-autoregressive model generates the outputs in parallel,
so the inference speed is faster, 
but the quality of generated result is relatively poor.

As shown in Table \ref{tab:intro}, PaddleSpeech has implemented the following commonly used autoregressive acoustic models:
Tacotron 2 and Transformer TTS,
and non-autoregressive acoustic models: SpeedySpeech, FastPitch and FastSpeech 2.

\paragraph{Vocoder}
As shown in Table \ref{tab:intro}, PaddleSpeech has implemented the following vocoders:
WaveFlow, Parallel WaveGAN, MelGAN, Style MelGAN, 
Multi Band MelGAN, and HiFi GAN.

\paragraph{Implementation}
The PadddleSpeech TTS implementation of FastSpeech 2 adopts some improvement from
FastPitch and uses MFA to obtain the forced alignment (the original FastSpeech
paper uses Tacotron 2).
Notably, the speech feature parameters of the acoustic model and the vocoder of one TTS pipeline should be the same.
Detailed settings can be found in the sample config file\footnote{\url{https://github.com/PaddlePaddle/PaddleSpeech/blob/develop/examples/csmsc/tts2/conf/default.yaml}} on CSMSC dataset.


\begin{table}[t!]
\centering
\resizebox{\columnwidth}{!}{
\begin{tabular}{c|c c|c}
\toprule
    &   \textbf{ Acoustic Model} & \textbf{Vocoder} & \textbf{MOS} $\uparrow$ \\
\midrule
ESPnet & Fastspeech 2 & PWGAN & 2.55 $\pm$ 0.19 \\ \midrule
     & Tacotron 2  & PWGAN & 3.69 $\pm$ 0.11 \\ 
     & Speedyspeech & PWGAN & 3.79 $\pm$ 0.09 \\ 
     & Fastspeech 2 & PWGAN & 4.25 $\pm$ 0.09 \\
PaddleSpeech & Fastspeech 2 & Style MelGAN & 4.32 $\pm$ 0.10 \\
     & Fastspeech 2 & MB MelGAN & 4.43 $\pm$ 0.09 \\
     & Fastspeech 2 & HiFi GAN & 4.72 $\pm$ 0.08 \\
\midrule
\end{tabular}
}
\caption{The MOS evaluation with 95\% confidence intervals 
for TTS models trained using CSMSC dataset. PWGAN stands for
Parallel WaveGan, MB MelGAN stands for Multi-Band MelGAN.
}
\label{tb:tts}
\end{table}

\paragraph{Results}
We report the mean opinion score (MOS) for naturalness evaluation 
in Table \ref{tb:tts}.
We use the crowdMOS toolkit \cite{ribeiro2011crowdmos}, where 14 Mandarin samples (see Appendix \ref{app:ttsexample}) from these 7 different models were presented to 14 workers on Mechanical Turk.
As shown in Table \ref{tb:tts}, 
PaddleSpeech can largely outperform ESPnet on Mandarin TTS.
The main reason is that PaddleSpeech TTS has a better text frontend as shown in Table \ref{tab:preprocess}.
Compared with other models, Fastspeech 2 with HiFi GAN can achieve the best results.







\section{Conclusion}

This paper introduces PaddleSpeech, an open-source, easy-to-use, all-in-one speech processing toolkit.
We illustrated the main design philosophy behind this toolkit to conduct development and research on various speech-related tasks accessible.
A number of reproducible experiments and comparisons show that PaddleSpeech achieves state-of-the-art or competitive performance with the most popular models on standard benchmarks.

\section{Acknowledgment}

We sincerely thank the anonymous reviewers for their valuable comments and suggestions. 
This work was supported by the National Key Research and Development Project of China (2020AAA0103503).

\bibliography{main.bib}
\bibliographystyle{acl_natbib}

\clearpage

\appendix


\section{TTS Examples}
\label{app:ttsexample}

We use the following sentences as the MOS evaluation test set in Table \ref{tb:tts}.

\begin{itemize}
\item 早上好，今天是2020/10/29，最低温度是-3°C。
\item 你好，我的编号是37249，很高兴为您服务。		
\item 我们公司有37249个人。		
\item 我出生于2005年10月8日。		
\item 我们习惯在12:30吃中午饭。		
\item 只要有超过3/4的人投票同意，你就会成为我们的新班长。		
\item 我要买一只价值999.9元的手表。		
\item 我的手机号是18544139121，欢迎来电。		
\item 明天有62\%的概率降雨。		
\item 手表厂有五种好产品。		
\item 跑马场有五百匹很勇敢的千里马。		
\item 有一天，我看到了一栋楼，我顿感不妙，因为我看不清里面有没有人。
\item 史小姐拿着小雨伞去找她的老保姆了。		
\item 不要相信这个老奶奶说的话，她一点儿也不好。
\end{itemize}

\end{CJK}
\end{document}